\documentstyle[floats,epsf,twocolumn,prd,aps]{revtex}
\begin{document}

\draft
%
%
\newcommand{\nc}{\newcommand}
\nc{\bea}{\begin{eqnarray}}
\nc{\eea}{\end{eqnarray}}
\nc{\beq}{\begin{equation}}
\nc{\eeq}{\end{equation}}

\nc{\half}{\frac{1}{2}}
\nc{\EH}{{}^3{\rm H}}
\nc{\EHe}{{}^3{\rm He}}
\nc{\UHe}{{}^4{\rm He}}
\nc{\GLi}{{}^6{\rm Li}}
\nc{\ZLi}{{}^7{\rm Li}}
\nc{\ZBe}{{}^7{\rm Be}}
\nc{\DH}{{\rm D}/{\rm H}}
\nc{\EHeH}{^3{\rm He}/{\rm H}}
\nc{\GLiH}{{}^6{\rm Li}/{\rm H}}
\nc{\ZLiH}{{}^7{\rm Li}/{\rm H}}

\nc{\me}{m_{\rm e}}
\nc{\etal}{{\it et al.}}
\nc{\ve}[1]{{\bf #1}}

%
%

\nc{\AJ}[3]{{Astron.~J.\ }{{\bf #1}{, #2}{ (#3)}}}
\nc{\anap}[3]{{Astron.\ Astrophys.\ }{{\bf #1}{, #2}{ (#3)}}}
\nc{\ApJ}[3]{{Astrophys.~J.\ }{{\bf #1}{, #2}{ (#3)}}}
\nc{\apjs}[3]{{Astrophys.~J.,\ Suppl.\ Ser.\ }{{\bf #1}{, #2}{ (#3)}}}
\nc{\apjl}[3]{{Astrophys.~J.\ Lett.\ }{{\bf #1}{, L#2}{ (#3)}}}
\nc{\app}[3]{{Astropart.\ Phys.\ }{{\bf #1}{, #2}{ (#3)}}}
\nc{\araa}[3]{{Ann.\ Rev.\ Astron.\ Astrophys.\ }{{\bf #1}{, #2}{ (#3)}}}
\nc{\arns}[3]{{Ann.\ Rev.\ Nucl.\ Sci.\ }{{\bf #1}{, #2}{ (#3)}}}
\nc{\arnps}[3]{{Ann.\ Rev.\ Nucl.\ and Part.\ Sci.\ }{{\bf #1}{, #2}{ (#3)}}}
\nc{\MNRAS}[3]{{Mon.\ Not.\ R.\ Astron.\ Soc.\ }{{\bf #1}{, #2}{ (#3)}}}
\nc{\mpl}[3]{{Mod.\ Phys.\ Lett.\ }{{\bf #1}{, #2}{ (#3)}}}
\nc{\Nat}[3]{{Nature }{{\bf #1}{, #2}{ (#3)}}}
\nc{\ncim}[3]{{Nuov.\ Cim.\ }{{\bf #1}{, #2}{ (#3)}}}
\nc{\nast}[3]{{New Astronomy }{{\bf #1}{, #2}{ (#3)}}}
\nc{\np}[3]{{Nucl.\ Phys.\ }{{\bf #1}{, #2}{ (#3)}}}
\nc{\pr}[3]{{Phys.\ Rev.\ }{{\bf #1}{, #2}{ (#3)}}}
\nc{\PRC}[3]{{Phys.\ Rev.\ C\ }{{\bf #1}{, #2}{ (#3)}}}
\nc{\PRD}[3]{{Phys.\ Rev.\ D\ }{{\bf #1}{, #2}{ (#3)}}}
\nc{\PRL}[3]{{Phys.\ Rev.\ Lett.\ }{{\bf #1}{, #2}{ (#3)}}}
\nc{\PL}[3]{{Phys.\ Lett.\ }{{\bf #1}{, #2}{ (#3)}}}
\nc{\prep}[3]{{Phys.\ Rep.\ }{{\bf #1}{, #2}{ (#3)}}}
\nc{\RMP}[3]{{Rev.\ Mod.\ Phys.\ }{{\bf #1}{, #2}{ (#3)}}}
\nc{\rpp}[3]{{Rep.\ Prog.\ Phys.\ }{{\bf #1}{, #2}{ (#3)}}}
\nc{\ibid}[3]{{\it ibid.\ }{{\bf #1}{, #2}{ (#3)}}}

\wideabs{
\title{Determination of $\Omega_b$ From Big Bang Nucleosynthesis 
       in the Presence of Regions of Antimatter}

\author{Elina Sihvola \cite{maile}}
\address{Department of Physics, University of Helsinki,
         P.O.Box 9, FIN-00014 University of Helsinki, Finland}


\maketitle

\begin{abstract}

Production of regions of antimatter in the early universe is predicted in many 
baryogenesis models. Small scale antimatter regions would annihilate during or 
soon after nucleosynthesis, affecting the abundances of the light elements. 
In this paper we study how the acceptable range in $\Omega_b$ changes in 
the presence of antimatter regions, as compared to the standard big bang 
nucleosynthesis.   
It turns out that it is possible to produce at the same time both a low
$\UHe$ value ($Y_p<0.240$) and a low D/H value ($\DH<4\times10^{-5}$), 
but overproduction of $\ZLi$ is unavoidable at large $\Omega_b$.

\end{abstract}

\pacs{PACS numbers: 26.35.+c, 25.43.+t, 98.80.Cq, 98.80.Ft}
}

%
%


\section{Introduction}

Big bang nucleosynthesis (BBN) in the presence of regions of antimatter 
in a baryo-asymmetric
universe has been studied  recently in several papers 
\cite{RehmPRL,KSPRL,KSPRD,RehmPRD},
first by Rehm and Jedamzik \cite{RehmPRL}.
Production of antimatter domains is predicted in many baryogenesis models. 
If the universe is baryo-asymmetric and the domain structure is of the right 
distance scale, antimatter would get annihilated during or soon after 
nucleosynthesis, affecting the abundances of the light elements. 

Annihilation after or shortly before recombination is strongly constrained 
by the cosmic microwave background (CMB) and cosmic diffuse gamma (CDG) 
radiation. If the antimatter regions are small enough to annihilate well 
before recombination, but after weak freeze-out, the strongest constraint 
on the amount of antimatter is obtained from big bang nucleosynthesis. 
The relevant scales are $10^{-5}-1$ pc.

In two earlier papers \cite{KSPRL,KSPRD} we derived constraints on the antimatter
fraction allowed by BBN observations. We fixed the mean 
baryon density to $\eta=6\times10^{-10}$, corresponding to $\Omega_bh^2=0.022$. 
In this paper we extend the study to a range of values of $\eta$. 
We search for values of $\eta$ that would be forbidden in the standard big bang 
nucleosynthesis (SBBN),  but allowed in the presence of antimatter (antimatter 
big bang nucleosynthesis, ABBN).


\section{Antimatter nucleosynthesis}

\subsection{Initial Conditions}

We consider a situation where the early universe contains regions of antimatter, 
surrounded by a background of ordinary matter.

We model an antimatter region by a sphere with initially constant 
antibaryon density $n_b$. The antimatter sphere is surrounded by a 
spherical matter shell with equal baryon density. 
The initial structure is set by three parameters: 
(1) the antimatter fraction $R=f_v/(1-f_v)$, where $f_v$ is the volume fraction 
of the antimatter region,  
(2) the net baryon density $\bar n_b=n_b(1-R)/(1+R)$, and 
(3) the physical radius of the antimatter region, $r_{\rm A}$. 
The distances are given in comoving units in meters at $T=$ 1 keV. 
One meter at $T=$ 1 keV corresponds to $4.26\times10^6$ m today. 
The net baryon density $\bar n_b$ is given as the baryon-to-photon ratio
$\eta$ today, which is related to $\Omega_b$ by 
$\eta=274\times10^{-10}\Omega_bh^2$.

\subsection{Lithium}

We have studied BBN with antimatter regions in two previous papers 
\cite{KSPRL,KSPRD}. 
We refer the reader to these papers for the details on physics and computations,  
and present in this paper only the new effects added since.

In our earlier papers we did not focus on lithium, 
whose observational limits are less secure than those of deuterium and helium.
In this paper we consider also the production of $\GLi$ and $\ZLi$.

The SBBN yield of $\GLi$ is small, $\GLiH\sim10^{-14}$. 
In the presence of antimatter additional $\GLi$ is produced by 
photodisintegration and annihilation
of $\ZLi$ and $\ZBe$ and by non-thermal nuclear reactions. 

About a half of the energy released in annihilation is carried away 
by photons and electrons, about a half by neutrinos, 
and a small fraction by nuclear debris.
Energy release in the form of photons or electrons
leads to $\GLi$ production via the nonthermal reactions
$\UHe+\EH\rightarrow\GLi+n$ and $\UHe+\EHe\rightarrow\GLi+p$,
as first noted by Jedamzik \cite{JedLi6}. 
The implications of this process on particle physics have been discussed
by Jedamzik \cite{JedLi6} and by Kawasaki \etal \cite{Kawasaki00}.

\begin{figure}[tbh]
\epsfysize=7.0cm
\smallskip
\epsffile{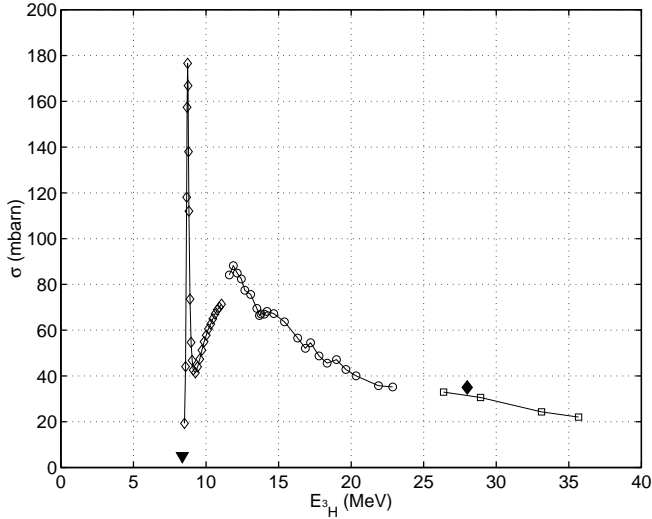}
\vspace{0.2cm}
\caption[a]{\protect 
  Cross section of the $\UHe(\EH,n)\GLi$ reaction, 
  calculated from the cross section of the inverse reaction.  
  The original $\GLi(n,\alpha)\EH$ data:  
    \cite{Overley74} (diamonds), 
    \cite{Bartle74} (circles), and 
    \cite{Bartle83} (squares).
  Not all the data points of the original papers have been included.
  The filled diamond presents the direct measurement by 
  Koepke and Brown \cite{Koepke}.
  The threshold (8.39 MeV) is indicated by a triangle.
} \label{fig:li6cross}
\end{figure}

Photons and electrons initiate electromagnetic cascades, 
which first lose energy rapidly through photon-photon 
pair production and inverse Compton scattering. When the photon energies fall 
below $E_c=\me^2/(80T)$, 
the photon-photon interactions cease and the 
thermalization continues much more slowly through pair production on a nucleus and
Compton scattering on electrons. 
During this phase the photons may also photodisintegrate nuclei. 

Photodisintegration of $\UHe$ leaves energetic $\EHe$ and $\EH$ ions, 
with energies up to $E_3=(E_c-E_{\rm tr})/4$, where $E_{\rm tr}=19.9(20.6)$ MeV is the
threshold of the photodisintegration reaction 
$\UHe+\gamma\rightarrow\EH(\EHe)+p(n)$.
These ions may further produce $\GLi$ via
reaction $\UHe+\EH(\EHe)\rightarrow\GLi+n(p)$.  
The threshold of this reaction in the laboratory frame  
is 8.39 MeV for $\EH$ and 7.05 MeV for $\EHe$. 
The photon energy must exceed 53.4(48.8) MeV in order to produce a
$\EH(\EHe)$ ion with energy above the threshold. 
The production of $\GLi$ begins when $E_c$ rises above 48.8 MeV,
i.e., when the temperature falls below $T=67$ eV.

We get a smaller $\GLi$ production from $\UHe$ photodisintegration than
Jedamzik \cite{JedLi6}, by a factor of 3.  The difference  
is mainly due to Compton scattering, which was ignored in 
\cite{JedLi6} in the thermalization of photons. 

Also $\EHe$ and $\EH$ ions from $\UHe+\bar p(\bar n)$
annihilation produce $\GLi$  via the same non-thermal reaction, 
as discussed by Rehm and Jedamzik \cite{RehmPRD}.   
The $\GLi$ ions produced  by this process remain
close to the annihilation zone, and are therefore  more likely annihilated later, 
than the $\GLi$ ions from $\UHe$  photodisintegration, which are distributed uniformly.
Still, the annihilation-generated $\EHe$ and $\EH$ dominate 
the production of $\GLi$.

We are aware of only one direct measurement on the cross section of the 
$\UHe+\EH\rightarrow\GLi+n$ reaction \cite{Koepke}. 
Fortunately, there are data on the inverse reaction
$\GLi+n\rightarrow\UHe+\EH$. 
We use the principle of detailed balance to calculate the
cross section of $\UHe+\EH\rightarrow\GLi+n$ from its inverse 
reaction.  The cross section is plotted in Fig. \ref{fig:li6cross}.
For the $\UHe+\EHe$ reaction we use the same cross section, 
shifted in energy according to the threshold energy. 

The photodisintegration of $\GLi$ into the channel $\UHe+p+n$ begins at 
$T=0.88$ keV, and that of  $\ZLi(\ZBe)$ at $T=$1.3(2.1) keV. 

We have implemented into our code all the the $\GLi$ production processes described above.  
The slowing down of ions, the cascade spectrum, and the spectra of the annihilation debris  
are as in \cite{KSPRD}.
The photodisintegration reactions included in our code, together with references 
to the cross-section data, are given in Table I. We use 
the total single photoneutron cross section $\ZLi(\gamma,n)X$ \cite{ADNT38} 
for the $\ZLi(\gamma,n)\GLi$ reaction. 
In principle this leads to an overestimation of the $\GLi$ production, since the reaction
$\ZLi(\gamma,n)X$ may include other reaction channels. This is, however, unimportant, 
since the production of $\GLi$ is dominated by other processes.

\begin{figure}[tbh]
\epsfysize=7.0cm
\smallskip
\epsffile{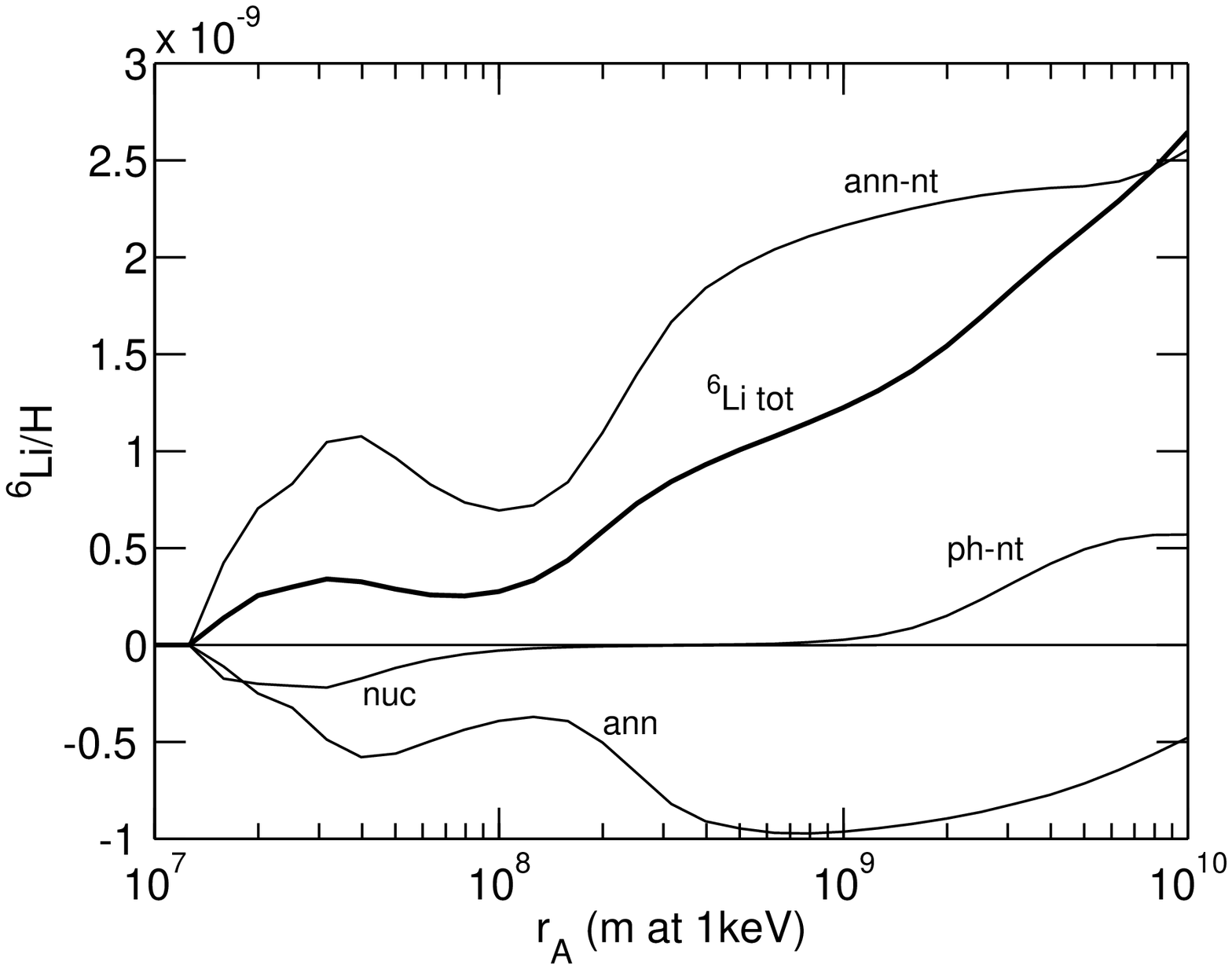}
\vspace{0.2cm}
\caption[a]{\protect 
  The $\GLi$ yield and major contributions to it,
  for $R=0.01$ and $\eta=6\times10^{-10}$. 
  Shown are the total $\GLi$ yield ('tot'), 
  production via reaction $\UHe+\EH(\EHe)\rightarrow\GLi+n(p)$
  by $\EH$ and $\EHe$ ions from annihilation ('ann-nt') and 
  photodisintegration ('ph-nt') of $\UHe$. 
  The negative lines show the destruction of $\GLi$ by
  annihilation ('ann') and thermal $\GLi+n$ ('nuc'). 
}
\label{fig:li6contrib}
\end{figure}

Figure \ref{fig:li6contrib} shows the $\GLi$ yield and major contributions 
to it, as a function of the antimatter radius, for $R=0.01$ 
and $\eta=6\times10^{-10}$. The $\GLi$ production is dominated by nonthermal 
reaction on $\UHe$ by $\EH$ and $\EHe$ produced in $\UHe+\bar p(\bar n)$
annihilation.  
The production via $\UHe$ photodisintegration becomes important at scales 
larger than $10^9$ m.
At scales $r_{\rm A}<10^8$ m (annihilation at $T>10$ keV) $\GLi$ is destroyed via 
the thermal neutron reaction $\GLi+n\rightarrow\UHe+\EH$.

The rise in the $\GLi$ production at scales $10^7$--$10^8$ m is due to 
the inefficient slowing down of ions at high temperatures, when the 
thermal velocities of the plasma electrons exceed the velocity of the ion 
(see \cite{KSPRD}). The probability of a nuclear reaction is then higher.
Direct production of $\GLi$ through photodisintegration and annihilation of
$\ZLi$ are negligible effects, both contribute $\sim 0.1$ \% 
of the total $\GLi$.

\begin{figure}[tbh]
\epsfysize=7.0cm
\smallskip
\epsffile{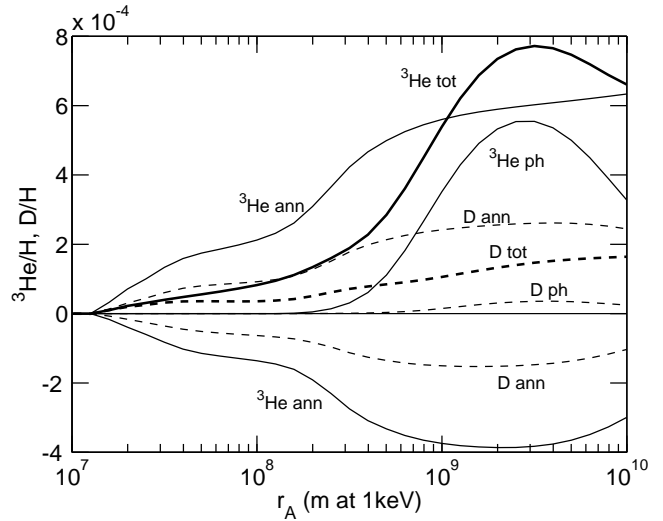}
\vspace{0.2cm}
\caption[a]{\protect 
  Major contributions to the production of $\EHe$ (solid lines) 
  and D (dashed lines), 
  for $R=0.01$ and $\eta=6\times10^{-10}$. 
  Shown are the total yield ('tot') from which the SBBN yield 
  has been subtracted, 
  production through annihilation ('ann') and 
  photodisintegration ('ph') of $\UHe$, and destruction
  by annihilation ('ann').
}
\label{fig:he3contrib}
\end{figure}

For comparison, the yields of $\EHe$ (including $\EH$) and D are shown 
in Fig. \ref{fig:he3contrib}.
Production of D is dominated by $\UHe+\bar p(\bar n)$ annihilation. 
For $\EHe$ production, both annihilation and photodisintegration of $\UHe$ 
are important. Spallation of $\UHe$ by energetic neutrons contributes less 
than 10\% of the D yield and less than 5\% of the $\EHe$ yield, 
and has been omitted from the figure. Photodisintegration is responsible for 0.5\% of
the destruction of $\EHe$ and for 2\% of the destruction of D.

\begin{table}
\begin{tabular}{lrl}
Reaction & Threshold & Ref. \\ \hline
$\GLi+\gamma \rightarrow \UHe+p+n$  & 3.70 MeV & \cite{ADNT38}   \\
$\GLi+\gamma \rightarrow \UHe$+D    & 1.48 MeV & \cite{Robertson81},
                                                 not included    \\
$\ZLi+\gamma \rightarrow \GLi+n$    & 7.25 MeV & \cite{ADNT38}   \\
$\ZBe+\gamma \rightarrow \GLi+p$    & 5.61 MeV &                 \\
$\ZLi+\gamma \rightarrow \UHe+\EH$  & 2.47 MeV & \cite{Skopik79} \\
$\ZBe+\gamma \rightarrow \UHe+\EHe$ & 1.59 MeV &                 \\
$\ZLi+\gamma \rightarrow \UHe+2n+p$ &10.96 MeV & \cite{ADNT38}   \\
$\ZBe+\gamma \rightarrow \UHe+2p+n$ & 9.31 MeV &                 \\
\end{tabular}
\caption[a]{\protect
  Photodisintegration channels for nuclei with $A=6,7$, with threshold energies 
  and references to the cross-section data.
  For $\ZBe$ we use the $\ZLi$ data, shifted in energy according 
  to the threshold energy.
  Photodisintegration of nuclei with $A\le4$ is as in \cite{KSPRD}.
}
\end{table}

\subsection{Diffusion}

The mixing of matter and antimatter proceeds through 
diffusion and collective hydrodynamic flow of the plasma. 
As different diffusion constants appear in the literature, 
we present here the constants used in our code.

\begin{table*}
\begin{tabular}{lllll}
Scattering process  & 
Diffusion constant  & 
cross section       &
Eq.                 &
statistics          \\ \hline
neutron-electron & 
   $D_{\rm ne} = (3/8)(\sqrt{(\pi T/2\me)}/(\sigma_{\rm ne}n_{\rm e}))
                 (K_2(z)/K_{2.5}(z))$             &
   $\sigma_{\rm ne} = 8\times10^{-4}$ mbarn     &
   \ref{rel_dconst},\ref{mobility}              &
   MB                                          \\ 
neutron-proton & 
   $D_{\rm np} = (3/8)\sqrt{(\pi T/m_{\rm p})}
                 /(\sigma_{\rm np}n_{\rm p}) $  &
   $\sigma_{\rm np}$  as in \cite{AHS87}        &
   \ref{nonrel_dconst}                          &  
   non-rel.                                    \\ 
ion-electron & 
    $D_{\rm ie} = (3/4)(\me^2/(\pi\alpha^2Z_i^2\Lambda n_{\rm e}))
                   (K_2(z)e^z/(z^2+2z+2))$         &
   $\sigma_C(p) = 4\pi\alpha^2\Lambda E^2/p^4$.   &
   \ref{rel_dconst},\ref{mobility}               &
   MB                                           \\ 
ion-ion & 
   $D_{\rm ij} = (3/8)\sqrt{(\pi T)/(2\mu_{ij})}
                  /(\sigma_{\rm pp}Z_i^2Z_j^2n_j)$   &
   $\sigma_{\rm pp} = (4\pi\alpha^2\Lambda)/(9T^2)$  &
   \ref{nonrel_dconst}                               &  
   non-rel.                                          \\ 
electron-photon &  
   $D_{{\rm e}\gamma} = (3/4)T/(\sigma_{\rm T}\epsilon_\gamma)$ &
   $\sigma_{\rm T}=667$ mbarn                                   &
   \ref{rel_dconst},\ref{mobility}                              &
   FD                                                          \\ 
\end{tabular}
\caption[a]{\protect
  Diffusion constants. 
  $K_n(z)$ is the $n$th modified Bessel function with $z=\me/T$.
}
\end{table*}

We use the momentum transfer method \cite{Present} to calculate the diffusion constants. 
Consider the diffusion of a particle species $k$.  The partial pressure of 
component $k$ is balanced by the force associated with the momentum transfer with other
plasma components,
\beq
     -\nabla P_k + \sum_{j\ne k} \ve F_{kj} = 0.  \label{momtransfer}
\eeq

If both $k$ and $j$ are non-relativistic, the collisional force is given by
\beq
     \ve F_{kj} = n_kn_j \int\int d^3\ve u_k d^3\ve u_j 
              f_k(\ve u_k)f_j(\ve u_j)\vert\ve u_{jk}\vert
                   \sigma_{kj}^{\rm t} \mu \ve u_{jk}, \label{nonrelf}  
\eeq
where $f_k(\ve u)$, $f_j(\ve u)$ are the velocity distributions of the two particle species,
$\ve u_{jk}=\ve u_j-\ve u_k$ is the relative velocity, $\mu$ the reduced mass, 
and $\sigma_{kj}^{\rm t}$ the transport cross section in the center-of-mass frame.
Assuming a small deviation from the Maxwellian distribution for $k$, such that 
$\langle \ve u_k\rangle = \ve v_k$, we obtain
\beq
    \ve F_{kj} =  -n_kn_j\frac83 \left(\frac{2T\mu}{\pi}\right)^{1/2} 
                    \sigma_{kj}^{\rm t} \ve v_k.
\eeq
Putting together Eqs. (\ref{momtransfer}) and (\ref{nonrelf}), and the continuity equation, 
we end up with the diffusion equation, with a diffusion constant equal to
\beq
    D_{kj} = \frac38 \left( \frac{\pi T}{2\mu}\right)^{1/2}
                     \left( n_j\sigma_{kj}^{\rm t}\right)^{-1} 
               \qquad \hbox{(non-rel)}.   \label{nonrel_dconst}
\eeq

In the case of a heavy non-relativistic gas diffusing into a light relativistic gas  
the collisional force takes the form 
\beq
    \ve F_{kj} = -\frac1b n_k\ve v_k,
\eeq
leading to a diffusion constant
\beq
      D_{kj} = bT \qquad\hbox{(rel.)},   \label{rel_dconst}
\eeq
where the mobility $b$ is given by
\beq
      -\frac{1}{b}\ve v_k = 
         \int d^3\ve p \rho_j(\ve p) \frac {\vert\ve p\vert}E 
         \sigma_{kj}^{\rm t}\ve p.   \label{mobility}
\eeq
Here $\rho_j$ is the phase space density of particle
$j$ in a frame moving with velocity $\ve v_k$ with respect to the laboratory frame.

We give our diffusion coefficients in Table II.
In calculating the proton-electron diffusion coefficient, 
we have integrated over the
energy dependence of the Coulomb cross section 
$\sigma_C(p) = 4\pi\alpha^2\Lambda E^2/p^4$.

\subsection{Transfer of energy and momentum}

The annihilation of a nucleon-antinucleon pair produces a number of pions 
(on average 5) which decay into photons, muons and neutrinos. 
The muons decay further into electrons and neutrinos.
Electrons and photons initiate electromagnetic cascades.
If one of the annihilating pair is a nucleus, the annihilation 
leaves nuclear remnants with energies of the order of 10 MeV.  

We discussed the spreading of the annihilation products and 
the electromagnetic cascades in \cite{KSPRD}. 
We ignored the effects on the kinetic
behaviour of the plasma, as we estimated them to be small. 
Here we discuss these effects in more detail.

The annihilation products spreading out from the annihilation zone cause
an effective force that tends to push (anti)matter away from the annihilation zone.
The kinetic temperature of the plasma, however, 
is not affected by the energy deposit from
the annihilation products, as we will argue below. 

The universe is radiation dominated during the era we are interested in. 
The energy released in annihilation corresponds to a small fraction of the energy
in the photon background. 
The photons form an essentially homogeneous heat bath, 
which will efficiently dilute the
annihilation energy, if only the energy transfer between photons and plasma particles is
rapid enough. To see if that is the case, we must compare the energy transfer rate with the
annihilation time scale.

\begin{figure}[tbh]
\epsfysize=6.8cm
\smallskip
\epsffile{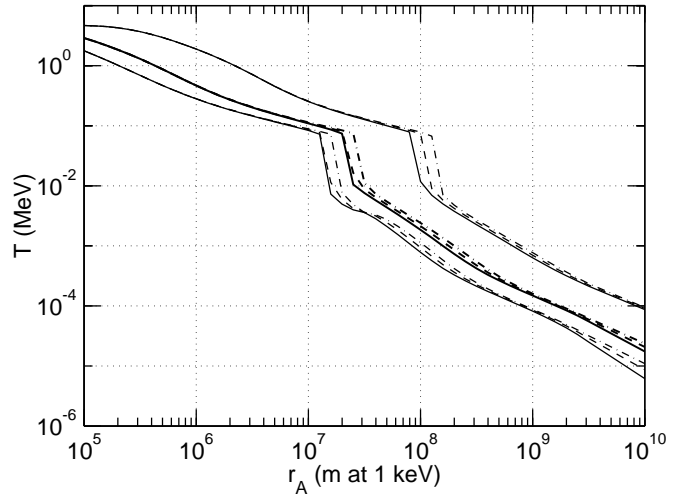}
\vspace{0.2cm}
\caption[a]{\protect 
  Annihilation temperature. Shown is the temperature at which 
  10\% (up), 50\% (mid), and 90\% (down) of antimatter has been annihilated,
  for $R=0.01$ and 
  $\eta_{10}\equiv 10^{10}\eta=9$ (solid), 
  $\eta_{10}=6$ (dashed) and
  $\eta_{10}=4$ (dash-dotted).
  Before nucleosynthesis the mixing of matter and antimatter proceeds 
  through diffusion of free neutrons.
  If annihilation is not complete before the onset of nucleosynthesis 
  ($T\approx80$ keV), it is delayed until proton diffusion becomes effective.
}
\label{fig:anntemp}
\end{figure}

In Fig. \ref{fig:anntemp} we plot the annihilation temperature versus the
radius of the antimatter region. During the annihilation phase the temperature of the
universe typically falls by one decade.

Electrons transfer energy to photons through inverse Compton scattering or, 
equivalently, through Thomson drag.
The relaxation time for the electron temperature in the photon bath is
given by
\beq
     t_{\rm eq} = \frac38 \frac{\me}{\sigma_{\rm T}\epsilon_\gamma}
                \approx 1.1\times10^{-7}{\rm s}\left(\frac{T}{\rm keV}\right)^{-4} . 
\eeq
This is small compared with the annihilation time scale,
implying that electrons and photons maintain the same temperature. 

In the annihilation on a nucleus, a small fraction of the energy 
is carried away by fragments of the nucleus. The ions are slowed down by Coulomb 
collisions, first with electrons and at lower energies with ions. 
Scattering on ions dominates the slowing down of an ion below energy
$E\sim16T$. 

Energy transfer between electrons and ions is inefficient due to their mass 
difference. When energy is injected into the plasma, a thermal energy distribution 
is first established within electrons and ions separately, possibly at different kinetic 
temperatures \cite{Spitzer}.  The thermalization between ions and electrons then 
occurs with a relaxation time
\beq
    t_{\rm eq} = \frac3{8\sqrt{2\pi}} \frac{A_im_{\rm p}\me}
                  {n_{\rm e}Z_i^2\alpha^2\Lambda}
                  \left( \frac{T_{\rm e}}{\me}\right)^\frac32
               \approx 0.2{\rm s}\frac {A_i}{\eta_{10}}
                  \left( \frac{T_{\rm e}}{\rm keV} \right)^{-\frac32} .
\eeq
This also is small compared with the annihilation time scale,
which implies that ions keep
the same temperature with electrons and photons.  
We may conclude that the effect of the annihilation on the plasma
temperature can be neglected.

The annihilation energy may still distort the photon spectrum, leaving an observable
remnant in the CMB.  The absence of observed distortion constrains the annihilation 
occurring below 1 keV.

Although the nuclear fragments carry only a small fraction of the annihilation 
energy, they carry a large fraction of the momentum. 
The momentum is transferred into the plasma near the annihilation zone via 
collisions with plasma particles. 
This causes a force that resists the flow of matter into the annihilation zone. 
The effective pressure gradient along the radial direction is
\beq
   \nabla P = \int\int f(E,\theta) F(E) \cos\theta d\Omega dE  \label{ionpress}.
\eeq
Here $f(E,\theta)$ is the spectral density of ions flowing into direction $\theta$, 
$F(E)=dE/dx$ is the energy loss per unit distance due to collisions, and $\theta$ 
is the angle between the ion's velocity and the radial direction.

This effect was implemented into our computer code, but the effect turned out to be
negligible. At the largest the effect was to reduce by 1\% the temperature 
at which the annihilation is complete.


\section{Results}

\subsection{Production of isotopes in ABBN}

\begin{figure*}[tph]
\hbox{
\epsfysize=7.0cm
\epsffile{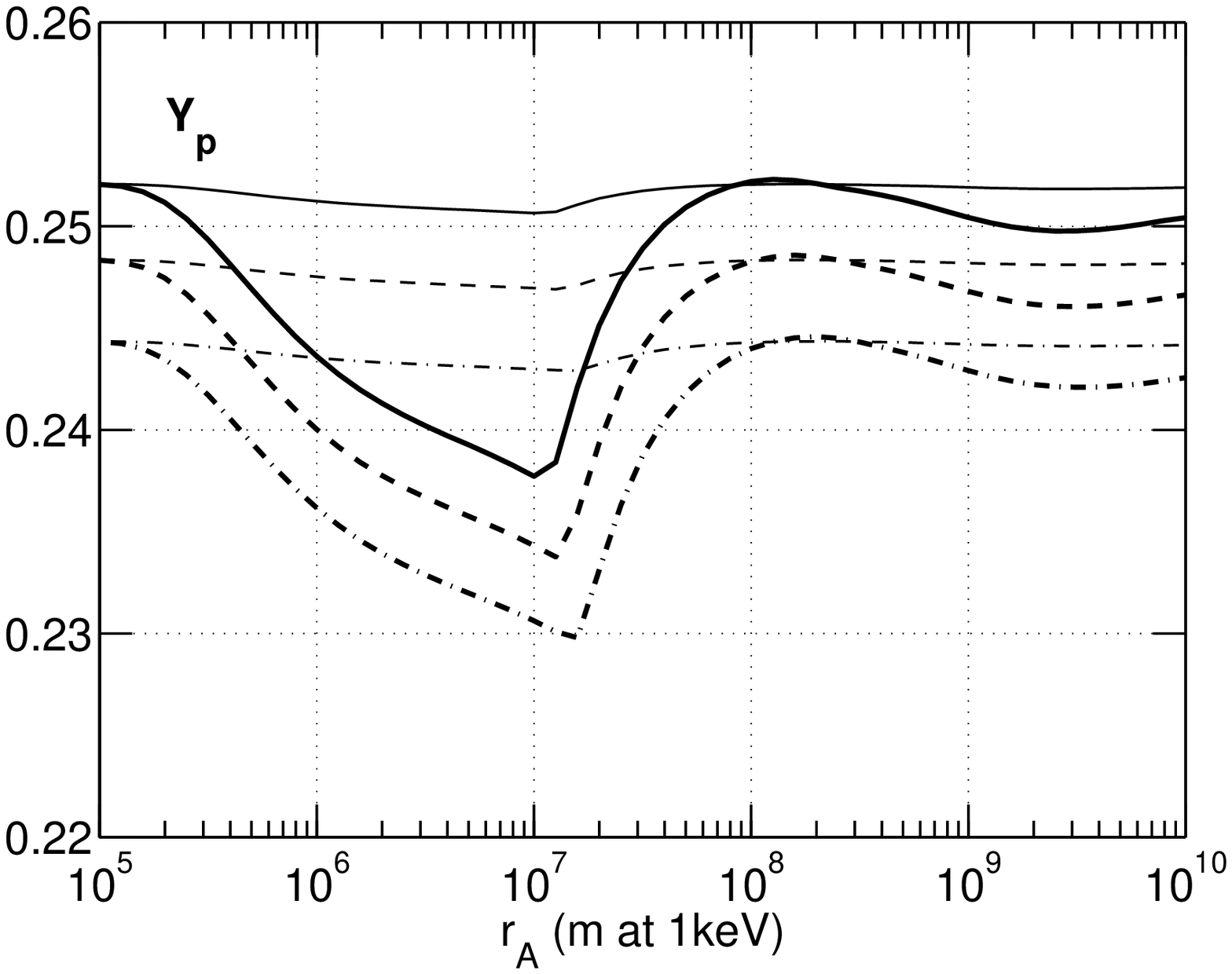}
\epsfysize=7.0cm
\epsffile{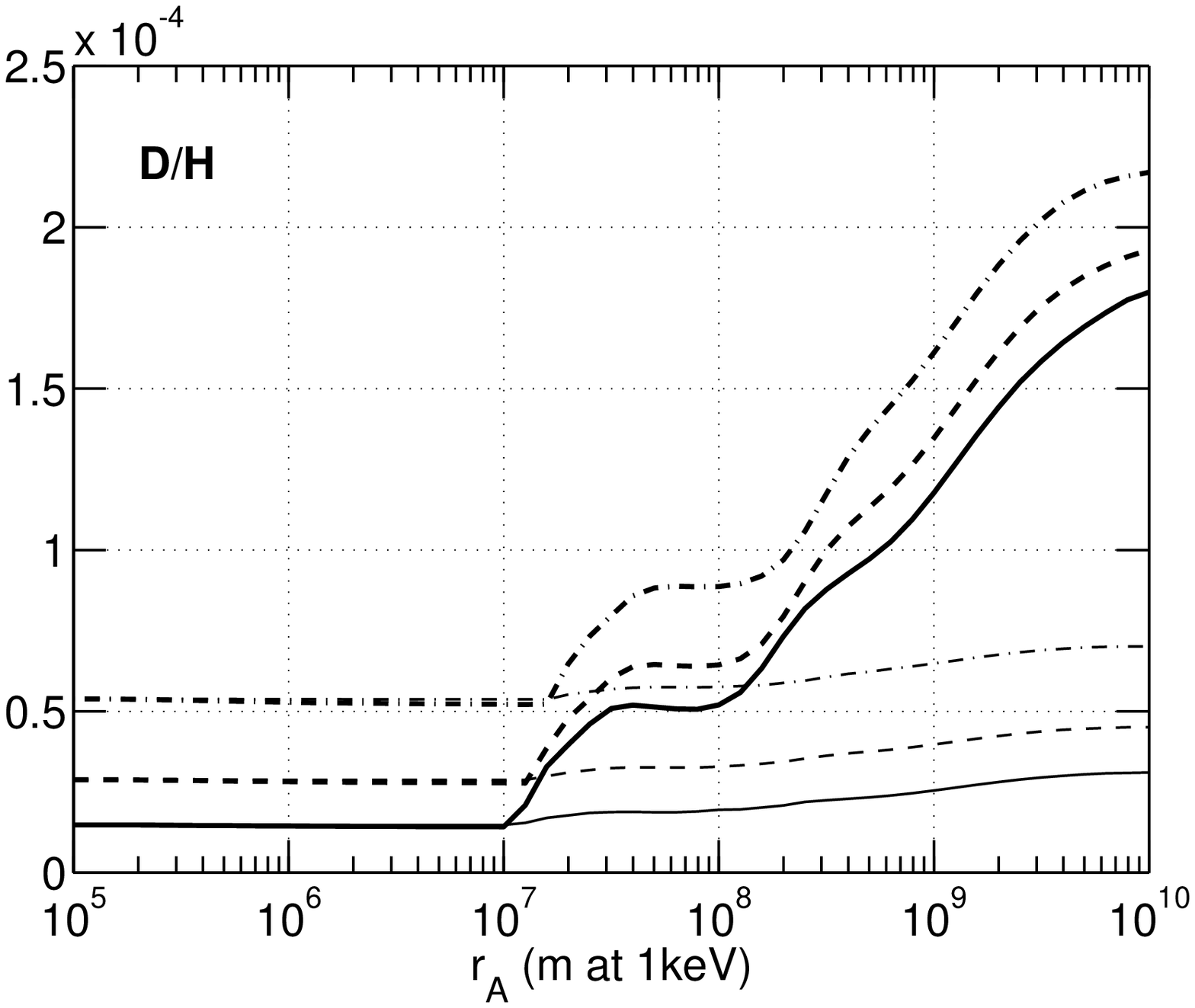}
}
\hbox{
\epsfysize=7.0cm
\epsffile{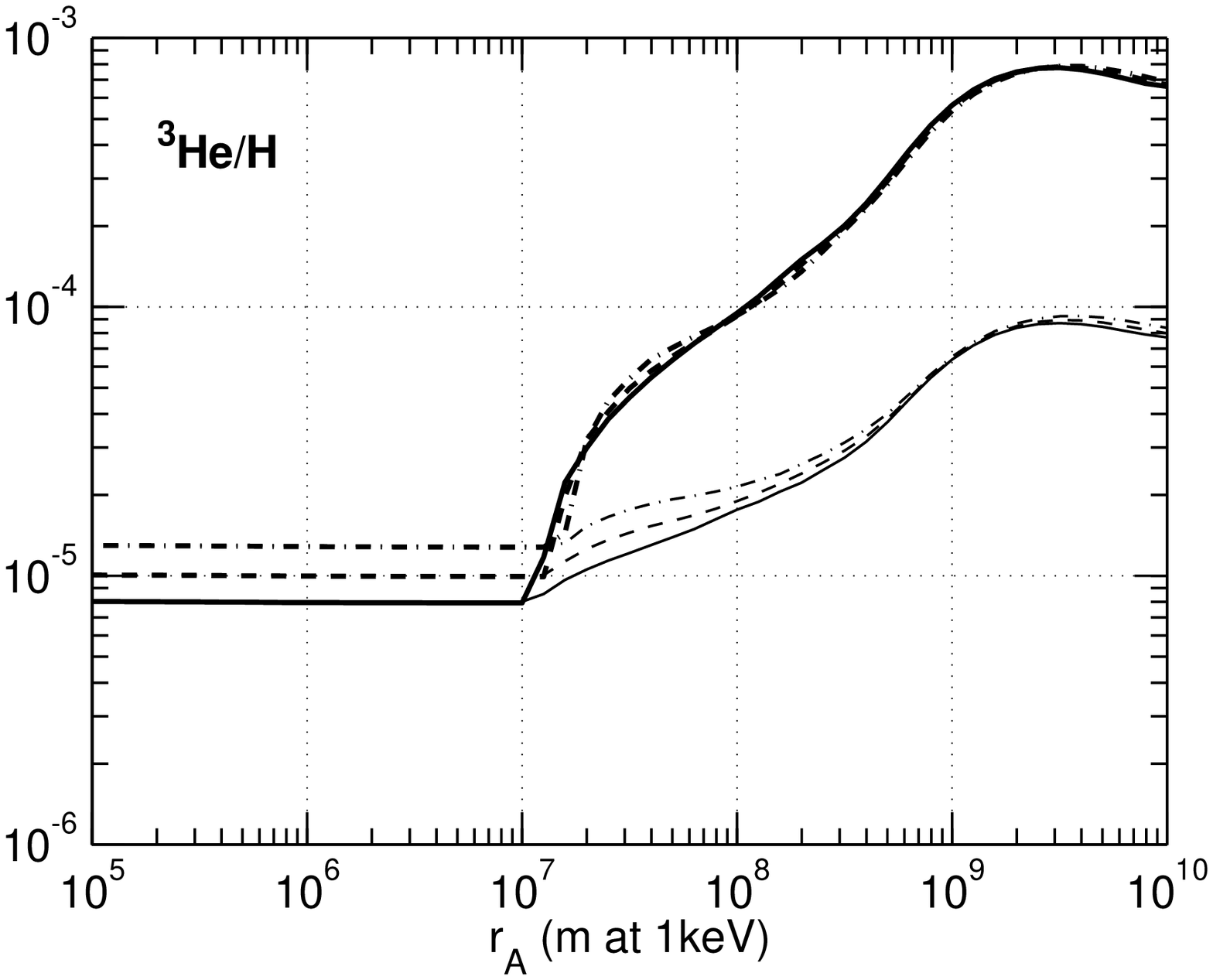}
\epsfysize=7.0cm
\epsffile{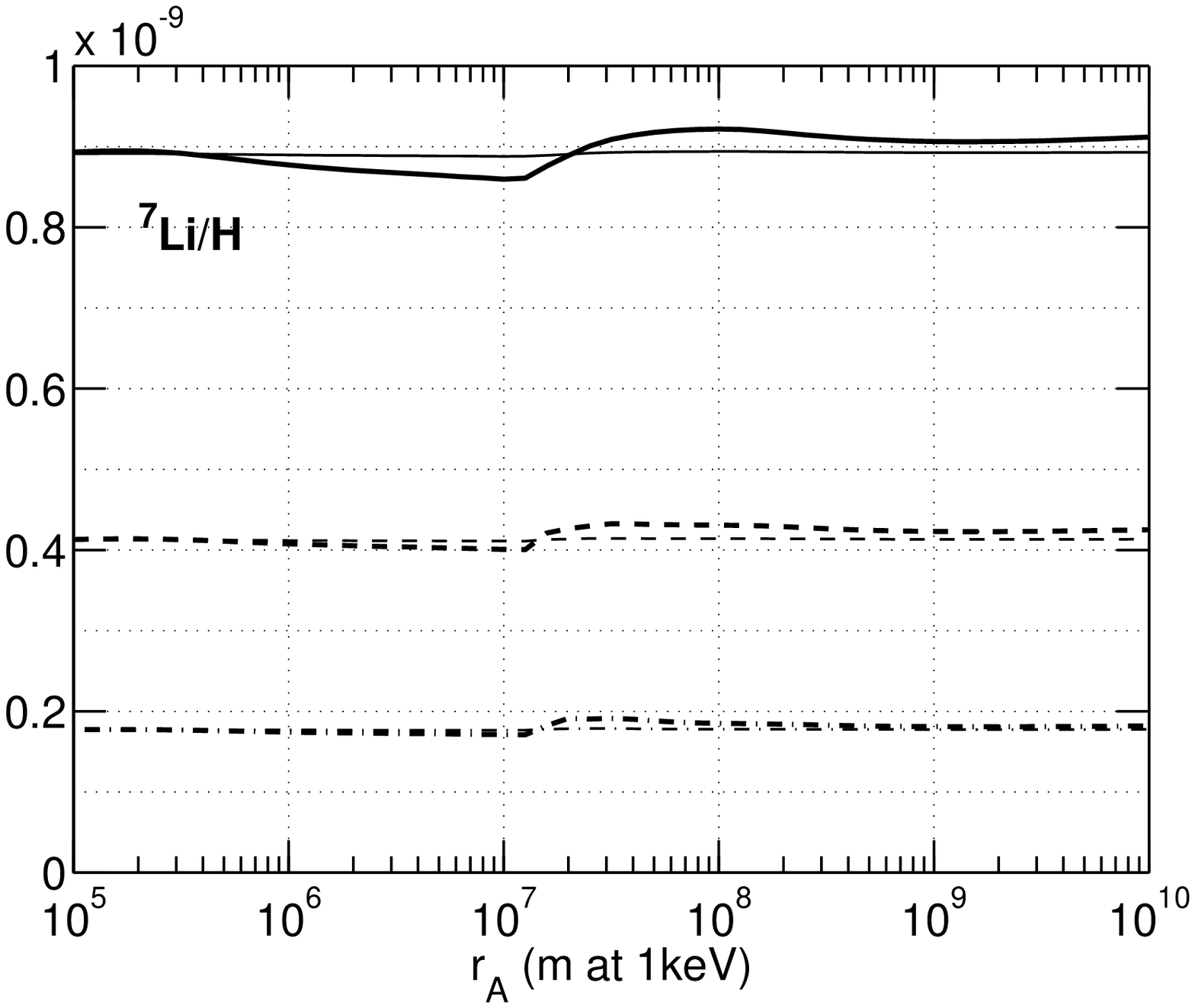}
}
\epsfysize=7.0cm
\epsffile{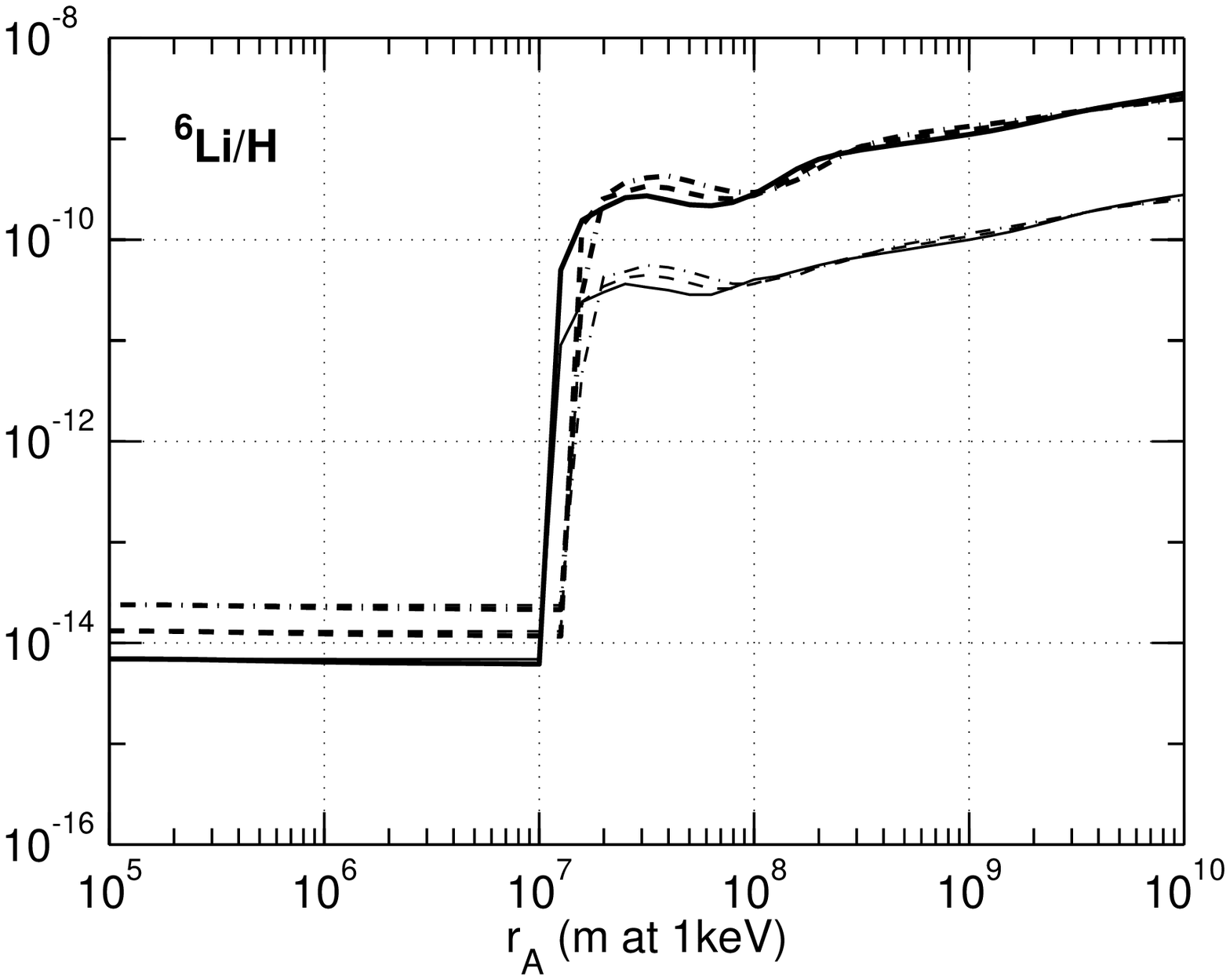}
\vspace{0.2cm}
\caption[a]{\protect 
  Isotope yields, versus the radius of the antimatter region, for
   $\eta=4\times10^{-10}$ (dot-dashed lines),
   $\eta=6\times10^{-10}$ (dashed lines), and
   $\eta=9\times10^{-10}$ (solid lines). 
  The thick lines are for antimatter fraction $R=0.01$, 
  the thin lines for $R=0.001$. 
  On the left the results converge towards the SBBN yield.
} \label{fig:eta469}
\end{figure*}

Figure \ref{fig:eta469} shows the isotope yields for 
$\eta_{10}\equiv10^{10}\eta=4,6,9$ and for antimatter fractions 
$R=0.01$ and $R=0.001$.

The $\UHe$ yield is generally reduced in antimatter nucleosynthesis as compared 
to SBBN.
At small scales the reduction is due to annihilation of neutrons 
before nucleosynthesis, 
at large scales due to annihilation and photodestruction of $\UHe$. 
The relative reduction in $\UHe$ is rather independent of $\eta$.
Maximal reduction in $\UHe$ is found at scale $r_{\rm A}\approx10^{7}$ m.

Annihilation and photodisintegration of $\UHe$ produce $\EHe$ 
and D. 
The final $\EHe$ and D yields are a sum of the SBBN yield and production 
due to annihilation. The latter depends
only weakly on $\eta$. 

The $\ZLi$ abundance is only weakly affected by annihilation. 
The small increase in the $\ZLi$ yield at large scales is due to the ordinary
inhomogeneity effect: nucleosynthesis takes place at a local density $n_b$ 
higher than the net baryon density $\bar n_b$.

The $\GLi$ yield increases in ABBN, as discussed in Section II. 
Like in the case of D and $\EHe$, the final $\GLi$
yield is a sum of the SBBN yield and additional production due to annihilation, 
the latter being only weakly dependent of $\eta$. 

The yields of D, $\EHe$, $\GLi$, and $\ZLi$ are reduced together with 
that of $\UHe$ at scales around $10^7$ m.
The decrease in isotopes other than $\UHe$ is insignificant, 
except for large antimatter fractions, for which $\UHe$ is severely underproduced.
It is not possible in ABBN to significantly reduce D, $\EHe$, $\GLi$, or $\ZLi$, 
and simultaneously produce an acceptable $\UHe$ value.

\subsection{Observational constraints}

There are two competing estimates for the primordial helium abundance, 
both based on measurements on the $\UHe$ abundance in extragalactic 
low-metallicity HII regions. 
One group obtains
$Y_p = 0.234\pm0.003$ \cite{loHe} (``low $\UHe$''), the other group
$Y_p = 0.244\pm0.002$ \cite{hiHe} (``high $\UHe$'').
These values correspond, respectively, to 
$1.2<\eta_{10}<2.7$ and 
$2.7<\eta_{10}<5.8$ ($2\sigma$ limits) in SBBN. 
Peimbert {\etal }  have recently measured $Y = 0.2405\pm0.0018$ 
in the Small Magellanic Cloud, and based on this, estimate 
$Y_p = 0.2345\pm0.0026$ \cite{Peimbert}.

Burles and Tytler \cite{Tytler} claim to have established the
primordial deuterium abundance as $\DH = (3.3\pm0.25)\times10^{-5}$, 
based on a detected low deuterium abundance in three Lyman absorbtion systems.
Their D/H value gives $5.2<\eta_{10}<5.8$ in SBBN.  
O'Meara \etal \cite{OMeara} have recently added a fourth deuterium detection, 
and revised the combined estimate to $\DH = (3.0\pm0.4)\times10^{-5}$ 
($5.4<\eta_{10}<6.4$). 

A recent review by Steigman \cite{Steigman} suggests observational
constraints $0.228< Y_p <0.248$ and $2.9\times10^{-5}<\DH<4.0\times10^{-5}$.  
We adopt the $\UHe$ constraints and the upper D limit of Steigman, but we extend 
the lower D limit to $\DH>2.2\times10^{-5}$ to take into account the result of 
O'Meara {\etal} These constraints lead to the SBBN estimates
$1.2<\eta_{10}<5.8$ from $Y_p$ and 
$4.8<\eta_{10}<7.1$ from D/H.
We discuss also the implications of the low $\UHe$ estimate.

The best estimate of the primordial lithium abundance is obtained from the 
``Spite plateau'' \cite{Spite}. 
Bonifacio and Molaro \cite{BoMo97} obtain a present abundance
log$_{10}(\ZLiH) = -9.80\pm0.012\pm0.05$ for the Spite plateau. 
While a number of authors \cite{lowLi7} argue against 
significant depletion from the primordial abundance,
Pinsonneault \etal \cite{Pins} estimate a depletion factor of 0.2--0.4 dex.
To be conservative, we adopt the relaxed constraint
$\ZLiH < 4\times10^{-10}$ (SBBN $1.1<\eta_{10}<5.9$). 

A number of measurements on the $\GLi/\ZLi$ ratio in halo and disk stars 
\cite{li6obs} give results ranging from zero up to $\GLi/\ZLi<0.13$.
Based on this, Jedamzik adopts the tentative limit 
$\GLiH<7\times10^{-12}$ \cite{JedLi6} for the primordial $\GLi$. 
As $\GLi$ is very fragile and may have been affected by stellar processing, 
it cannot be ruled out that the primordial $\GLi$ is significantly higher.  
We choose to use only the $\UHe$, D, and $\ZLi$ constraints
in determining the allowed range in $\eta$.

\subsection{Fitting $\eta$}

As we have seen, in ABBN the $\UHe$ yield is reduced 
and the D yield increased with respect to SBBN. 
This indicates that we may find consistency with observations at a
high baryon density which is not allowed in SBBN.

\begin{figure}[tph]
\epsfysize=7.0cm
\smallskip
\epsffile{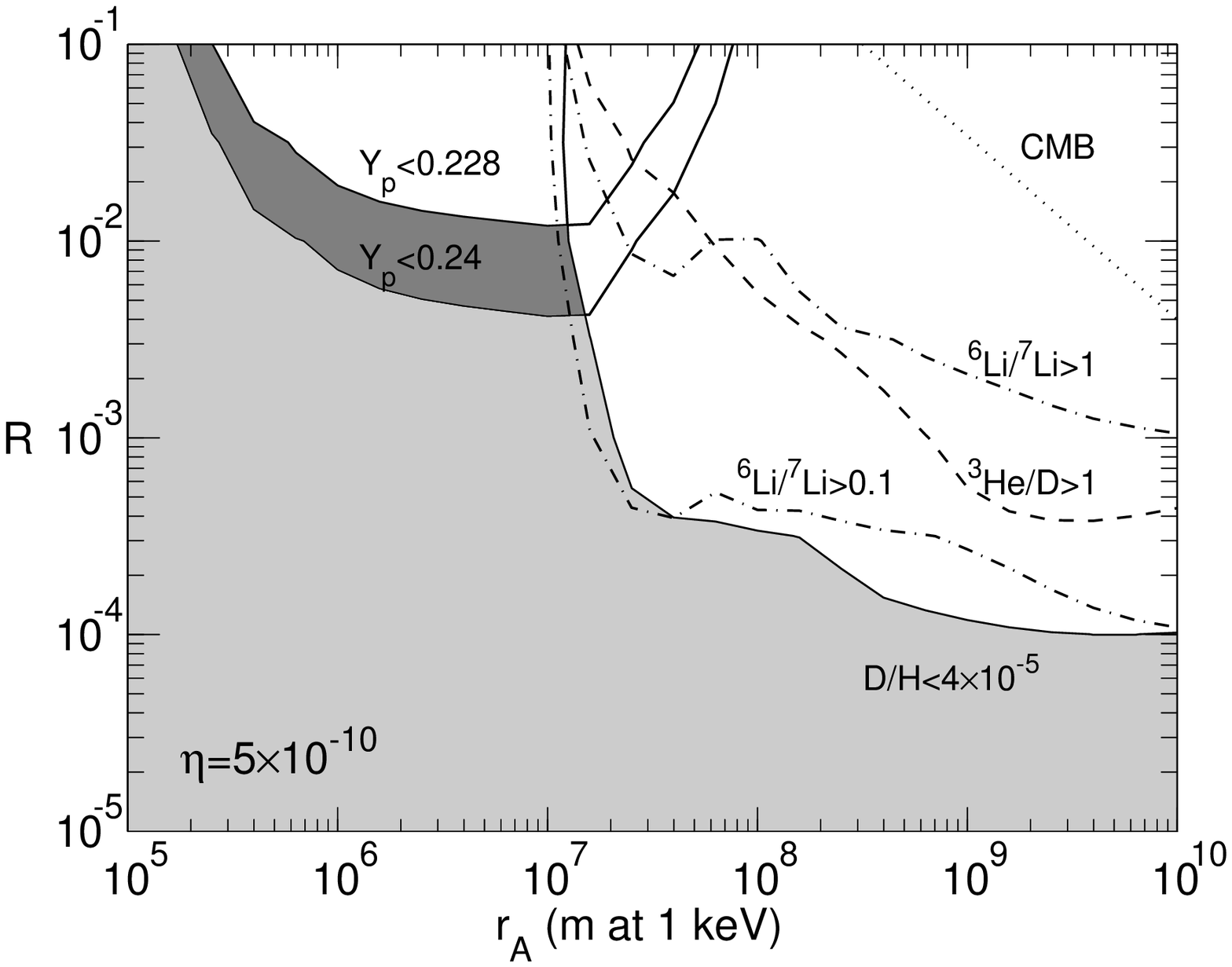}
\vspace{0.2cm}
\caption[a]{\protect 
  Observational BBN constraints on the\break  ($r_{\rm A},R$) plane, 
  together with the CMB constraint,\break for $\eta=5\times10^{-10}$.
  The lower left corner of the plot corresponds to SBBN.
  The shadowing indicates regions that satisfy the D and $\UHe$ constraints 
  $2.2\times10^{-5}<\DH<4.0\times10^{-5}$, $Y_p>0.228$, 
  and $Y_p<0.240$ (dark shading) or $Y_p<0.248$ (light shading). 
  The $\ZLi$ yield\break ($\ZLiH=2.8\times10^{-10}$) 
  is nearly constant over the parameter plane.
  For comparison we show also the constraint $\EHe/{\rm D}<1$ (dashed line)
  and the contours $\GLi/\ZLi=0.1$, $\GLi/\ZLi=1$ (dash-dotted lines). 

  The SBBN yields are 
    $\DH   = 3.8\times10^{-5}$,
    $\EHeH = 1.1\times10^{-5}$,
    $Y_p   = 0.2466$,
    $\GLiH = 1.7\times10^{-14}$, and 
    $\ZLiH = 2.8\times10^{-10}$.  
}
\label{fig:eta5}
\end{figure}

\begin{figure}[tph]
\epsfysize=7.0cm
\smallskip
\epsffile{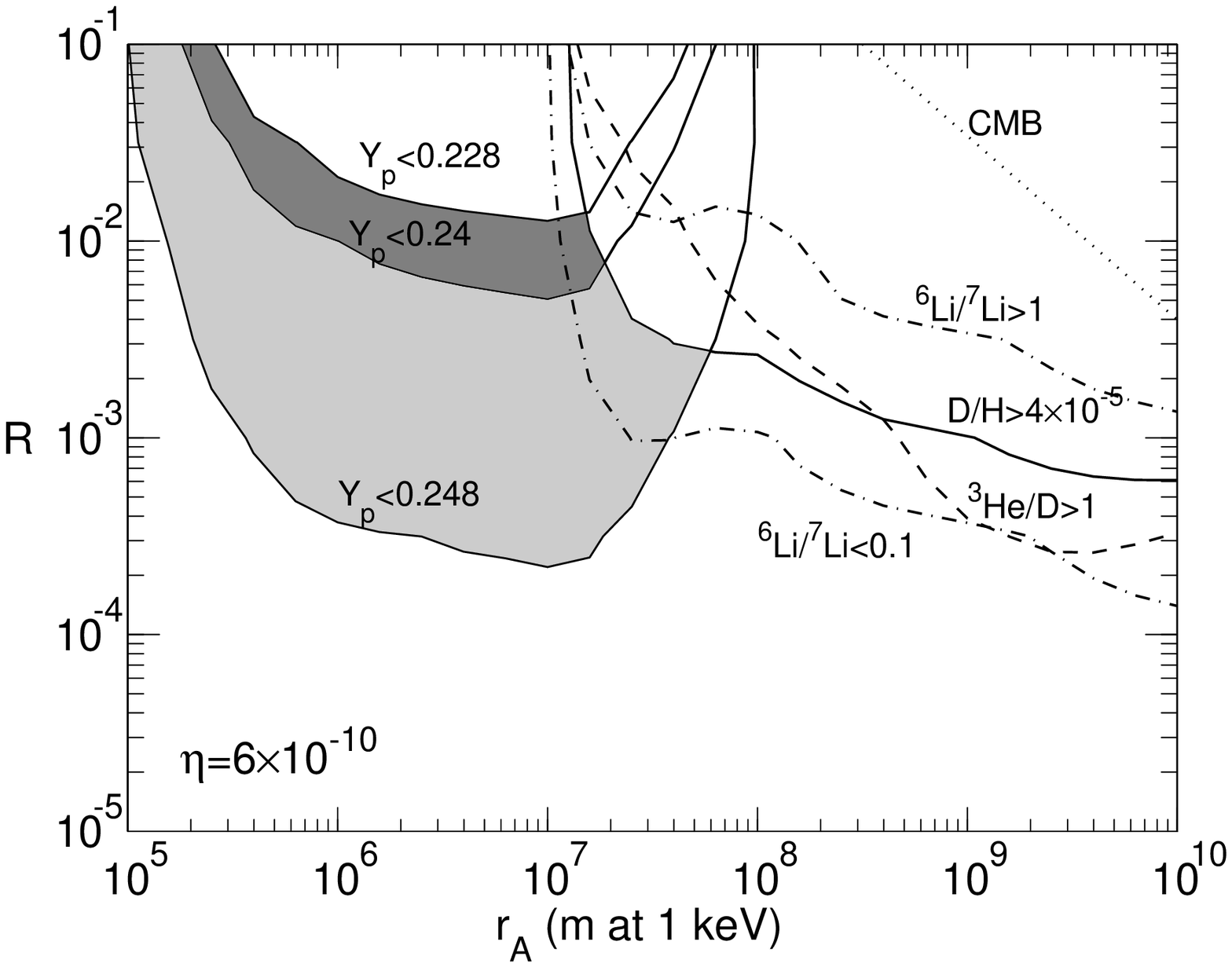}
\vspace{0.2cm}
\caption[a]{\protect 
  Same as Fig. \ref{fig:eta5}, but for $\eta=6\times10^{-10}$.
  The $\ZLi$ yield is $\ZLiH=4.1\times10^{-10}$.
  The SBBN yields are
    $\DH   = 2.9\times10^{-5}$,
    $\EHeH = 1.0\times10^{-5}$,
    $Y_p   = 0.2483$,
    $\GLiH = 1.3\times10^{-14}$, and 
    $\ZLiH = 4.1\times10^{-10}$.  
}
\label{fig:eta6}
\end{figure}

\begin{figure}[tph]
\epsfysize=7.0cm
\smallskip
\epsffile{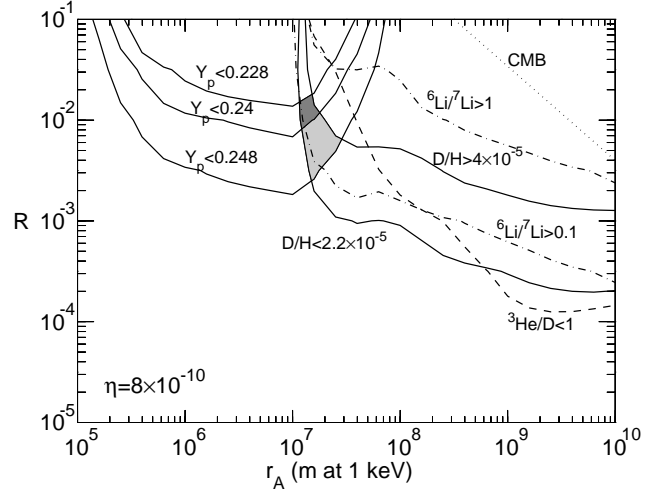}
\vspace{0.   cm}
\caption[a]{\protect 
  Same as Fig. \ref{fig:eta5}, but for $\eta=8\times10^{-10}$.
  The D and $\UHe$ constraints are satisfied in the shadowed region. 
  However, this high a baryon density is
  excluded by overproduction of $\ZLi$, whose yield $\ZLiH=7.3\times10^{-10}$ 
  is constant over the plane.
  The SBBN yields are
    $\DH   = 1.8\times10^{-5}$,
    $\EHeH = 8.6\times10^{-6}$,
    $Y_p   = 0.2510$,
    $\GLiH = 8.4\times10^{-15}$, and 
    $\ZLiH = 7.3\times10^{-10}$. 
}
\label{fig:eta8}
\end{figure}

Figures \ref{fig:eta5}-\ref{fig:eta8} show the observational constraints on the
($r_{\rm A},R$) plane, for different values of $\eta$.
Shown are the constraints $2.2\times10^{-5}<\DH<4.0\times10^{-5}$ 
and $0.228<Y_p<0.248$, $Y_p<0.240$.
For comparison we show also the constraint $\EHe/{\rm D}<1$ used in \cite{KSPRD}
and the contours $\GLi/\ZLi=0.1$ and $\GLi/\ZLi=1$.
The yield of $\ZLi$ is nearly constant over the plane, and is therefore not shown as contours.
The CMB constraint is obtained by demanding that the energy released in annihilation must not
exceed a fraction of $6\times10^{-5}$ of the CMB energy.

Figure \ref{fig:eta5} shows our results for $\eta=5\times10^{-10}$
($\Omega_bh^2=0.0182$).  The value
$\eta_{10}=5$ is consistent with observations in SBBN,  if we accept the high $\UHe$ limit,
$Y_p<0.248$.  The low $\UHe$ limit, $Y_p<0.240$, is inconsistent with the deuterium yield in SBBN, 
but in ABBN we find a consistency region at $R>0.004$, 
at scales $r_{\rm A}<10^7$ m. The constraints $Y_p<0.248$ and $\DH>2.2\times10^{-5}$ 
are satisfied everywhere in the parameter plane.

In Fig. \ref{fig:eta6} we show our results for 
$\eta=6\times10^{-10}$ ($\Omega_bh^2=0.0219$). 
This value is excluded in SBBN by overproduction of $\UHe$ (SBBN yield $Y_p=0.2483$). 
Lithium-7 ($\ZLiH=4.1\times10^{-10}$) is slightly above its upper limit.
In ABBN there is a region in the parameter space, where both $\UHe$ and D 
are well in their observed ranges. 
The required antimatter fraction depends on which 
$Y_p$ limit we adopt.  We find $R>2\times10^{-4}$ for $Y_p<0.248$  
and $R>0.005$ for $Y_p<0.240$.  
The $\ZLi$ yield, however, remains close to its SBBN value, and is only marginally
acceptable.

Figure \ref{fig:eta8} shows our results for 
$\eta=8\times10^{-10}$\break ($\Omega_bh^2=0.0292$). The $\ZLi$ yield $\ZLiH=7.3\times10^{-10}$ 
is clearly inconsistent with the present estimations of the primordial $\ZLi$ abundance.  Still, it
is interesting to note that it is possible to bring both D and $\UHe$ into their observed ranges. In
the region allowed by the D\break and $\UHe$ constraints the $\GLi/\ZLi$ ratio varies in the range
$0.06-0.4$.

Overproduction of deuterium sets a lower limit to the baryon density which can be 
accomodated in ABBN.  Since ABBN cannot reduce the D/H yield from 
the SBBN yield, without simultaneously severely underproducing
$\UHe$, a low baryon density that is excluded in SBBN by D/H overproduction 
is excluded in ABBN as well. With our adopted observational constraints 
($\DH<4.0\times10^{-5}$) this gives a lower limit $\eta>4.8\times10^{-10}$, 
corresponding to $\Omega_bh^2>0.0177$.

An upper limit to $\eta$ is obtained from overproduction of $\ZLi$, 
whose yield in ABBN is not significantly deviated from SBBN. 
With $\ZLiH<4\times10^{-10}$ we get 
$\eta<5.9\times10^{-10}$ ($\Omega_bh^2<0.0216$).
Thus the range allowed for $\eta$ by our adopted
observational constraints changes only slightly, from
$4.8 < \eta_{10} < 5.8$ in SBBN to
$4.8 < \eta_{10} < 5.9$ in ABBN.

Recent balloon experiments on CMB anisotropy, Boomerang and Maxima-1,  
suggest a large baryon density 
$\Omega_bh^2\sim0.030$, $\eta\sim8\times10^{-10}$ \cite{boommax} 
in clear conflict with SBBN. 
Antimatter nucleosynthesis alone is not a solution for this discrepancy, 
because of the overproduction of lithium at large $\eta$.
It can, however, help with D and $\UHe$.

In \cite{KSPRL,KSPRD} we placed upper limits on the antimatter fraction at 
$\eta=6\times10^{-10}$.  Our new results show that the limit 
is not sensitive to the value of $\eta$.


\section{Conclusions}

Nucleosynthesis in the presence of antimatter can produce simultaneously 
a small $\UHe$ mass fraction and a high $\DH$ ratio, allowing for a higher baryon
density than SBBN. Antimatter nucleosynthesis provides one possible solution 
to the discrepancy  between the low primordial $\UHe$ value suggested by some 
observations and the low D/H value.

The allowed baryon density is limited from above by $\ZLi$ overproduction.
The $\ZLi$ yield is almost unaffected by the presence of antimatter. 
A baryon density larger than $\Omega_bh^2=0.0216$ leads to a high lithium yield
$\ZLiH>4\times10^{-10}$, in conflict with the present estimates 
of the primordial value.

Since the $\ZLi$ yield is almost unaffected by ABBN, the upper limit to $\Omega_b$ 
is not much changed from SBBN. However, if one disregards the $\ZLi$ limit, 
ABBN allows larger values of $\Omega_b$ than SBBN, including the high value 
suggested by Boomerang and Maxima-1.


\section*{Acknowledgements}

I thank H. Kurki-Suonio for useful discussions and the 
Center for Scientific Computing 
(Finland) for computational resources.


\end{document}